\documentclass[aps,prl,showpacs,twocolumn,superscriptaddress,nofootinbib]{revtex4-1}  
\usepackage{graphicx,color}  
\usepackage{dcolumn}   
\usepackage{bm}        
\usepackage{amssymb} 
\usepackage{amsmath}
 \newcommand{\beq}{\begin{equation}}
\newcommand{\eeq}{\end{equation}}
\newcommand{\beqa}{\begin{eqnarray}}
\newcommand{\eeqa}{\end{eqnarray}}
\newcommand{\be}{\begin{equation}}
\newcommand{\ee}{\end{equation}}
\newcommand{\bea}{\begin{eqnarray}}
\newcommand{\eea}{\end{eqnarray}}

\begin{document}
\def\eg{{\it e.g.}}
\newcommand{\nc}{\newcommand}
\nc{\rnc}{\renewcommand}
\rnc{\d}{\mathrm{d}}
\nc{\D}{\partial}
\nc{\K}{\kappa}
\nc{\bK}{\bar{\K}}
\nc{\bN}{\bar{N}}
\nc{\bq}{\bar{q}}
\nc{\vbq}{\vec{\bar{q}}}
\nc{\g}{\gamma}
\nc{\lrarrow}{\leftrightarrow}
\nc{\rg}{\sqrt{g}}
\rnc{\[}{\begin{equation}}
\rnc{\]}{\end{equation}}
\nc{\nn}{\nonumber}
\rnc{\(}{\left(}
\rnc{\)}{\right)}
\nc{\q}{\vec{q}}
\nc{\x}{\vec{x}}
\rnc{\a}{\hat{a}}
\nc{\ep}{\epsilon}
\nc{\tto}{\rightarrow}
\rnc{\inf}{\infty}
\rnc{\Re}{\mathrm{Re}}
\rnc{\Im}{\mathrm{Im}}
\nc{\z}{\zeta}
\nc{\mA}{\mathcal{A}}
\nc{\mB}{\mathcal{B}}
\nc{\mC}{\mathcal{C}}
\nc{\mD}{\mathcal{D}}
\nc{\mN}{\mathcal{N}}
\rnc{\H}{\mathcal{H}}
\rnc{\L}{\mathcal{L}}
\nc{\<}{\langle}
\rnc{\>}{\rangle}
\nc{\fnl}{f_{NL}}
\nc{\gnl}{g_{NL}}
\nc{\fnleq}{f_{NL}^{equil.}}
\nc{\fnlloc}{f_{NL}^{local}}
\nc{\vphi}{\varphi}
\nc{\Lie}{\pounds}
\nc{\half}{\frac{1}{2}}
\nc{\bOmega}{\bar{\Omega}}
\nc{\bLambda}{\bar{\Lambda}}
\nc{\dN}{\delta N}
\nc{\gYM}{g_{\mathrm{YM}}}
\nc{\geff}{g_{\mathrm{eff}}}
\nc{\tr}{\mathrm{tr}}
\nc{\oa}{\stackrel{\leftrightarrow}}
\nc{\IR}{{\rm IR}}
\nc{\UV}{{\rm UV}}

\title{ An $SU(15)$ Approach to $B$ Anomalies} 

\author{Claudio Corian\`o and Paul H. Frampton}
\affiliation{Dipartimento di Matematica e Fisica, Universit\`{a} del Salento \\ and \\ INFN Sezione di Lecce, Via Arnesano, 73100 Lecce, Italy}

\date{\today}

\begin{abstract}
The most efficient way to extend the Standard Model of particle theory so that it
accommodates the observed B anomalies is the hypothesis
of a leptoquark which is a color-triplet 
weak-singlet boson with electric charge $Q=+\frac{2}{3}$.
We point out that SU(15) grand unification with a scalar leptoquark
in the adjoint representation gives a natural explanation for the violation of lepton flavour universality. 
We briefly discuss biquarks and bileptons.

\end{abstract}

\maketitle

The Standard Model of particle theory has remained robust
and only relatively recently have tantalising hints appeared from experiment 
about how
to extend it. When these hints have become more 
definite they
are likely to influence all of theoretical physics by clarifying
the choices which Nature has made. The Standard Model respects LFU = Lepton Flavour Universality, meaning that the three leptons
$ e, \mu, \tau$ have identical properties, in every way except for their different masses.
In B anomalies, we encounter experimental results
which violate LFU, and therefore are in disagreement with the standard model. For one recent
review of LHCb's results about LFU violation, see \cite{LHCb:2021trn}. 

\bigskip

\noindent
Muon-electron LFU is  explicitly violated by the present experimental data as follows.
For the neutral weak current, we have
\begin{equation}
R_K = \frac{ \Gamma(B^0 \rightarrow K^0 \mu^+\mu^-)}{\Gamma(B^0 \rightarrow K^0 e^+e^-)}
\label{RK}
\end{equation}
and the closely related ratio
\begin{equation}
R_{K^*} = \frac{ \Gamma(B^0 \rightarrow K^{*0} \mu^+\mu^-)}{\Gamma(B^0 \rightarrow K^{*0} e^+e^-)}
\label{RK*}
\end{equation}
Because the masses of $\mu^{\pm}$ and $e^{\pm}$ are negligible, the LFU property
of the standard model (SM) leads to predictions for the ratios in Eqs (\ref{RK},\ref{RK*}):
\begin{equation}
\left( R_K \right)_{SM} = \left( R_{K^*} \right)_{SM}  = 1.00 \pm 0.01
\label{predictions}
\end{equation}
The predictions in
Eq. (\ref{predictions}) depend only on the LFU property of the unadorned SM.\\
The latest results from LHCb and Belle II
can be summarised by:

\begin{equation}
R_K = 0.84 \pm 0.04
\label{RKexpt}
\end{equation}
and
\begin{equation}
R_{K^*} = 0.69 \pm 0.09
\label{RK*expt}
\end{equation}
which are at least $3\sigma$ below the SM values.
In neutral weak currents there are also measurements of the ratio

\begin{equation}
R_{K^+} = \frac{ \Gamma(B^+ \rightarrow K^{+} \mu^+\mu^-)}{\Gamma(B^+ \rightarrow K^{+} e^+e^-)}
\label{RKplus}
\end{equation}
with the result that
\begin{equation}
R_{K^+} =  0.74 \pm 0.09
\label{RKplusexpt}
\end{equation}
where the SM prediction from LFU is $R_{K^+} = 1.00\pm 0.01$.\\
Similar anomalies have been measured for the charged weak current
\begin{equation}
R_{D^{(*)}} = \frac{\Gamma(B\rightarrow D^{(*)} \tau^-\bar{\nu}_{\tau})}
{\Gamma(B\rightarrow D^{(*)}l^-\bar{\nu}_l)}
\label{RD}
\end{equation}
with $l = e, \mu$.
The experimental results are
\begin{equation}
R_{D}^{ratio} \equiv \frac {R_D^{expt}}{R_D^{SM}} = 1.29 \pm 0.17
\label{RDexpt}
\end{equation}
and
\begin{equation}
R_{D^*}^{ratio} \equiv \frac {R_{D^*}^{expt}}{R_{D^*}^{SM}} = 1.28 \pm 0.09
\label{RD*expt}
\end{equation}
which represent deviations from LFU of $1.7\sigma$ and $3.1\sigma$ respectively.

\bigskip

\noindent
One attempt at unification of strong and electroweak interactions\cite{Frampton:1989fu} involves the gauge group $SU(15)$
where all 15 states of a quark-lepton family are in the defining representation
and every possible leptoquark is present in the adjoint representation. 

\bigskip

\noindent
The adjoint has dimension 224 and contains 72 leptoquarks which transform
in irreducible representations of the standard model gauge group $(SU(3)_C, SU(2)_L)_Y$,  with $Q=T_3 + Y/2$ in four sets of 18 as follows.

\bigskip

\noindent
\underline{B = +1/3, L=+1}\\
$2(3, 2)_{-5/3}$ ~~~~ Q=(-1/3,-4/3) ~~~~ $ue^-,de^-$  \\
\noindent
$(3, 2)_{1/3}$ ~~~~~~~~   Q=(2/3. -1/3) ~~~~  $u\nu,d\nu $  

\bigskip
\bigskip

\noindent
\underline{B = -1/3, L=+1}\\
$2(3^*,1)_{-4/3}$ ~~~~~	Q=(-2/3) ~~~~~~~~$\bar{u}\nu$  \\
\noindent
$(3^*, 1)_{-10/3}$ ~~~~~Q=(-5/3) ~~~~~~~~~~~~~ $\bar{u}e^- $  \\
$(3^*, 3)_{-4/3}$ ~~~~~~	Q=(-5/3,-2/3,+1/3) ~~~~ $\bar{u}e^-,\bar{u}\nu,\bar{d}\nu$    \\

\noindent
\underline{B = +1/3, L=-1}\\
\noindent
$2(3, 1)_{4/3}$ ~~~~~	Q=(2/3) ~~~~~~~~~~~~~ $e^+d$    \\
\noindent
$(3, 1)_{10/3}$ ~~~~~ Q=(5/3)	~~~~~~~~~~~~~~ $e^+ d$  \\
\noindent
$(3, 3)_{4/3}$ ~~~~~~~ Q=(-1/3,2/3.5/3) ~~~~~ $\nu d,e^+d,e^+u$  \\

\noindent
\underline{B = -1/3, L=-1}\\
\noindent
$2(3^*, 2)_{5/3}$ ~~~~~	Q=(1/3,4/3) ~~~~~~~~~~~~ $e^+\bar{d},e^+\bar{u}$\\
\noindent
$(3^*, 2)_{-1/3}$ ~~~~~ Q=(-2/3, 1/3) ~~~~~~~~~~~  $\nu\bar{u},e^+\bar{u}$

\bigskip

\noindent
We see that the leptoquark favoured by $B$ anomalies occurs not only once but twice in the 224. It does
not appear in any of the smaller irreps 15, 105 or 120.

\bigskip

\noindent
The adjoint describes both the spin-one gauge bosons of SU(15) and a spin-zero Higgs
necessary \cite{Frampton:1990hz} for symmetry  breaking. Our first work is to choose between spin-one
and spin-zero for the leptoquark.
The spin-one hypothesis would imply that the leptoquark is a gauge boson of $SU(15)$.
In that case, if at least the first two families are treated sequentially as 15's, unless there
is an {\it ad hoc} assumption motivated by the data\cite{Cornella:2021sby}, muon-electron LFU will be
an inevitable consequence. 
Thus the leptoquark must be an adjoint Higgs scalar of $SU(15)$. In this case
it is natural that the $\mu$ Yukawa coupling be larger than the $e$ Yukawa
coupling by the ratio of their masses $M_{\mu}/ M_e \sim 200$, thus
explaining the observed violation of LFU discussed above.

\bigskip

\noindent
The one disadvantage of $SU(15)$, but only an aesthetic one and a stumbling
block we must initially ignore, is that
anomaly cancellation requires the addition of mirror fermions. An advantage
of $SU(15)$ is the absence of proton decay because all of the adjoint
components have well-defined $B$ and $L$ quantum numbers.
Smoking guns for $SU(15)$ include a predicted enhancement for 
$B \rightarrow K^{(*)}\nu \bar{\nu}$. Because of the lepton
mass dependence in the Higgs Yukawas, we predict significant
LFU-violating enhancements
relative to the SM for the decays $B^+\rightarrow K^+ \tau^+\tau^-$
and $B_s \rightarrow \tau+\tau^-$.
In an ingenious argument \cite{Glashow:2014iga}, it has been convincingly shown
that violation of LFU implies the occurrence of LFV decays which are
vanishing in the standard model. These will include the decays
$\tau\rightarrow\mu\gamma$, $\tau \rightarrow\mu\phi$ and 
$B_s \rightarrow \tau\mu$. The discovery of such LFV processes would 
lend support for the theory we have discussed.

\bigskip

\noindent
It will be exciting to learn from experiments about more
violations of LFU, as well as any examples of LFV. Such additional
input is necessary further to evolve the theory.\\
Because a leptoquark is suggested by the B anomalies and bileptons
are suggested by the 331-model,  we are tempted to believe that the
another type of bifermion, the biquark, appearing in the 224
of $SU(15)$ might also exist in Nature.
The 224 has 76 components with $B = L = 0$. The  remaining 148 include
the 72 leptoquarks listed above, 72 biquarks and 4 bileptons.\\
The 72 biquarks fall into two sets of 36:

\bigskip

\noindent
\underline{B = +2/3, L=0}\\

\noindent
$(3^*+6, 2)_{5/3}$ ~~~~~~~Q=(1/3,4/3) ~~~~~~~~ $ud,uu$   \\
\noindent
$(3^*+6, 2)_{-1/3}$ ~~~~~~ Q=(1/3,-2/3) ~~~~~~~ $ud, dd$   

\bigskip

\noindent
\underline{B = -2/3, L=0}\\

\noindent
$(3+6^*, 2)_{-5/3}$ ~~~~~~ Q=(-4/3,-1/3) ~~~~~~ $\bar{u}\bar{u},\bar{u}\bar{d}$\\
\noindent
$(3+ 6^*, 2)_{+1/3}$ ~~~~~~~~	Q=(-1/3,  2/3) ~~~~~~~~$\bar{u}\bar{d}, \bar{d}\bar{d}$

\bigskip

\noindent
Leptoquarks, as well as bileptons and biquarks, could represent a class of fundamental particles that, if discovered,  
 would represent new states in particle physics. Understanding the nature of such elementary particles, 
which should not be interpreted as composites, would be an important departure from the 
structure of the SM.

\bigskip

\noindent  
In the phenomenological analysis of tetraquarks (first discovered in 2003)
and pentaquarks (2015), the name "diquark" is used for two quarks behaving
together like a molecule, so a diquark is definitely a bound state and not an
elementary particle like a quark. At present the study of tetraquarks and
pentaquarks is successful \cite{Esposito:2021vhu} by using only diquarks without biquarks.
It will be interesting to discover whether biquarks become necessary in the
analyses. The distinction between diquark and biquark could be made
using the same criterion as used in \cite{Weinberg:1965zz} to decide whether
the deuteron is a bound state or elementary. 

\bigskip

\noindent
Finally, we discuss the four bileptons in the 224 which are in two
$SU(2)$ doublets $(Y^{--},Y^-)$ with $B=0, L=2$, and $(Y^{++},Y^+)$
with B=0, L=-2. In the context of the 331-model, they lead \cite{Frampton:1992wt} to the prediction
of a resonance in same-sign leptons at about $M_Y \simeq 1.3$TeV and
width $\Gamma_Y \simeq 0.05$ TeV.

\bigskip

\noindent
The bilepton resonance in $\mu^{\pm}\mu^{\pm}$ has been the subject of searches
by the ATLAS and CMS Collaborations at the LHC, starting in March 2021. In 
March 2022, ATLAS published an inconclusive result \cite{ATLAS:2022yzd} about the existence
of the resonance, putting only a lower mass limit $M_Y > 1.08$ TeV. CMS has
better momentum resolution and charge identification than ATLAS and should
be able to investigate the bilepton resonance proper.
Of the three classes of elementary bifermion (biquark, leptoquark, bilepton)
the one nearest to confirmation is the leptoquark which, as we have discussed,
is strongly indicated by the B anomalies.

\bigskip

\noindent
It seems possible that $SU(15)$ can act as an umbrella group which
unifies the SM and its 224 dimensional adjoint representation
does provide
a unique guide to the potential bifermions.

\end{document}